\documentclass[namedreferences]{SolarPhysics}
\usepackage{epsfig}

\newcommand{\pdiff}[2]{\frac{\partial #1}{\partial #2}}

\newcommand{\eqref}[1]{(\ref{eqn:#1})}
\newcommand{\expt}[1]{\,\textrm{e}^{#1}}
\newcommand{\s}{\,\textrm{s}}

\newcommand{\gauss}{\,\textrm{G}}
\newcommand{\mpsec}{\,\textrm{m}\,\textrm{s}^{-1}}
\newcommand{\kmsqpsec}{\,\textrm{km}^2\textrm{s}^{-1}}
\newcommand{\mxday}{\,\textrm{Mx}\,\textrm{day}^{-1}}
\newcommand{\mx}{\,\textrm{Mx}}

\begin{document}
\begin{article}
\begin{opening}

\title{Modelling the Global Solar Corona:} 
\subtitle{Filament Chirality Observations and Surface Simulations}

\author{A.R. \surname{Yeates}}
\author{D.H. \surname{Mackay}}
\institute{School of Mathematics and Statistics, University of St Andrews, Fife, KY16 9SS, Scotland \email{anthony@mcs.st-and.ac.uk}}
\author{A.A. \surname{van Ballegooijen}}
\institute{Harvard-Smithsonian Center for Astrophysics, 60 Garden Street, Cambridge, MA 02138, USA}

\runningauthor{A.R. Yeates \textit{et al.}}
\runningtitle{Modelling the Global Solar Corona}

\date{Received ; accepted }

\begin{abstract}

The hemispheric pattern of solar filaments is considered in the context of the global magnetic field of the solar corona. In recent work Mackay and van Ballegooijen have shown how, for a pair of interacting magnetic bipoles, the observed chirality pattern could be explained by the dominant range of bipole tilt angles and helicity in each hemisphere. This study aims to test this earlier result through a direct comparison between theory and observations, using newly-developed simulations of the actual surface and 3D coronal magnetic fields over a 6-month period, on a global scale.

In this paper we consider two key components of the study; firstly the observations of filament chirality for the sample of 255 filaments, and secondly our new simulations of the large-scale surface magnetic field. Based on a flux-transport model, these will be used as the lower boundary condition for the future 3D coronal simulations. Our technique differs significantly from those of other authors, where the coronal field is either assumed to be purely potential, or has to be reset back to potential every 27 days in order that the photospheric field remain accurate. In our case we ensure accuracy by the insertion of newly-emerging bipolar active regions, based on observed photospheric synoptic magnetograms. The large-scale surface field is shown to remain accurate over the 6-month period, without any resetting. This new technique will enable future simulations to consider the long-term build-up and transport of helicity and shear in the coronal magnetic field, over many months or years.

\end{abstract}

\end{opening}

\section{Introduction}
%----------------------------------------------------------------------------

Solar filaments are large regions of dense cool plasma suspended in the hot corona \cite{engvold1998}. They are well observed in H$\alpha$, either in absorption against the disk or in emission at the limb. For historical reasons in the latter case they are usually termed ``prominences''. Coronal magnetic fields are key to the existence of filaments \cite{mackay1}, as they both support the filament mass against gravity and help to insulate it from the surrounding corona.

\begin{figure}
\begin{center}
\leavevmode
\epsfig{file=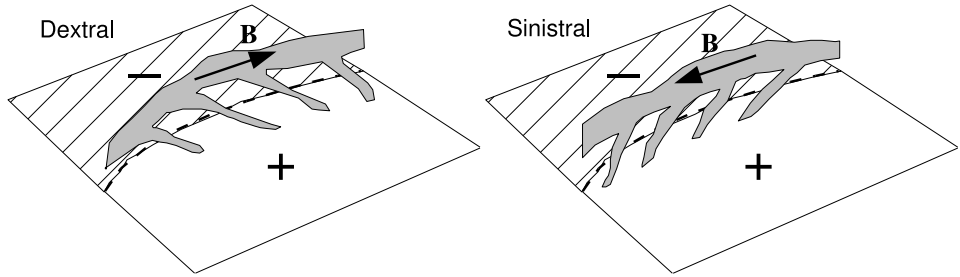,width=1.0\textwidth,clip=}
  \caption{Definition of filament chirality, showing the axial magnetic field ($\mathbf{B}$) direction, and barb orientation, in a dextral filament and a sinistral filament. The magnetic polarities in the photosphere beneath are also indicated.}
  \label{fig:chirality}
\end{center}
\end{figure}

Since the earliest magnetograph observations \cite{babcock1955} it has been known that filaments form above polarity inversion lines (PILs) in the photosphere, which divide regions of positive and negative magnetic flux. In the filament itself, the dominant magnetic field component is found to be along the filament's long axis \cite{foukal1971,leroy1989}. The filament may be classified as having either dextral or sinistral chirality (handedness) according to the direction of this dominant axial field \cite{martin1994}. Viewing the filament from the positive polarity side of the PIL, a right-pointing field is dextral and a left-pointing field is sinistral. This is illustrated in Figure \ref{fig:chirality}.

The property of filament chirality is notable because it follows a hemispheric pattern, whereby dextral filaments dominate in the Northern hemisphere and sinistral in the Southern hemisphere. The pattern was initially observed in polar crown prominences by \inlinecite{rust1967}, who noted that the field direction was opposite to that expected from differential rotation. It was later observed by \inlinecite{leroy1983} with Hanle-effect measurements, and then confirmed by \inlinecite{martin1994} and \inlinecite{pevtsov2003} using H$\alpha$ observations. The latter paper found that 80\% -- 85\% of quiescent filaments (which form between the remnants of decaying active regions) followed the rule. A feature of the hemispheric pattern is that it remains the same when the Sun's polar field reverses, approximately every 11 years.

\inlinecite{zirker1997} suggested that the hemispheric pattern could be produced by a combination of supergranular motions, differential rotation, and magnetic reconnection \cite{martens2001}, without recourse to unobservable subsurface phenomena as in some earlier theories \cite{vanballegooijen1990,rust1994,priest1996}. This suggests that the pattern for quiescent filaments could be a natural result of the observed evolution of bipolar active regions, and their interaction as they are transported poleward. A series of papers, beginning with \inlinecite{vanballegooijen1998}, have used coupled numerical modelling of the surface and coronal magnetic fields to investigate this idea. Having reached the conclusion that emerging active-region helicity does play an important role \cite{mackay2003}, \inlinecite{mackay3} gave a convincing explanation for the hemispheric filament pattern. However, this was a parameter study which considered only a pair of model magnetic bipoles in a localised region of the solar corona. In this study we aim to extend the model to the global corona, and to simulate the real Sun for a period of 6 months, allowing us to compare the local direction of the simulated magnetic field with the chirality of a large number of observed filaments during the period.

A distinction must be made between two approaches that have been used to model the coronal magnetic field based on photospheric flux transport. In the first approach the photospheric flux transport simulations are carried out independently from the coronal field modelling (\textit{e.g.} \opencite{mackay5b}; \opencite{wang2002}; \opencite{schrijver2003}). Here the flux transport simulations are used at discrete intervals, perhaps every 27 days, to supply the lower boundary condition for the extrapolation of a potential magnetic field in the coronal volume. The potential field is analysed at that time instant and then discarded, a new potential field being constructed after the next 27 days. In this approach, which has been successful in the past, there is evidently no direct coupling between the flux transport model evolving the photospheric field, and the extrapolated coronal fields. This precludes any memory in the coronal field of previous field line connectivity. 

In contrast, more recent flux transport models have coupled the photospheric and coronal magnetic fields, and include the models of \inlinecite{vanballegooijen1998} and \inlinecite{vanballegooijen2000}, developed to investigate the hemispheric pattern of filaments. The technique described in the latter paper has been used with small modifications in a series of papers (\opencite{mackay2}; \opencite{mackay2001}, \citeyear{mackay3}, \citeyear{mackay4}), and is used again in this study. As carried out in \inlinecite{mackay2}, the radial magnetic field component in the photosphere is taken as the initial boundary condition, and an initial potential coronal field extrapolated from it. Both the photospheric and coronal fields are then evolved together, allowing the build-up of magnetic energy and helicity in the coronal volume as a result of the photospheric motions.

\inlinecite{mackay2} applied the model to 10 filaments forming above an observed activity complex. In these simulations, the emergence of new flux was not included, so the photospheric field became less accurate over time. For this reason the photospheric fields were reset to the observed field after every 27 days of evolution. At this point, the sheared coronal field had to be discarded and a new potential field constructed. Although a significant advance on previous models with a purely potential coronal field, the build-up of magnetic energy and shear was limited to rather short periods of time compared to that on the real Sun. Despite the loss of accuracy in the surface field, the authors tried running the simulation for several months without resetting, and found in fact that the chirality results were slightly improved.

We have now modified the model, enabling us to simulate coronal magnetic fields for much longer periods of time without having to reset the photospheric or coronal fields every 27 days. Key to our new technique is the insertion of individual active regions throughout the simulation, based on the observed photospheric field. The new surface flux transport model is described and illustrated in this paper, simulating the real Sun over a particular 6-month period in 1999.

A sample of 255 filaments have been observed in H$\alpha$ images from Big Bear Solar Observatory (BBSO), over the same period. The chiralities of these filaments have been determined where possible, as described in this paper. In a follow-up paper the full 3D coronal field evolution will be computed, coupled to the new surface simulation. The resulting magnetic skew will be compared with the chirality of filaments at the locations where they were observed.

The structure of this paper is as  follows. Section \ref{sec:filaments} describes the filament data, and the observed chiralities. The numerical model for surface flux transport is outlined in Section \ref{sec:model}, while Section \ref{sec:flux} describes in detail the method for including newly-emerging active regions. Example surface simulation results are given in Section \ref{sec:results}, and a concluding summary in Section \ref{sec:conclusions}.

\section{Filament Observations} \label{sec:filaments}
%----------------------------------------------------------------------------

The filament data set consists of 255 filaments identified in H$\alpha$ images from Big Bear Solar Observatory, over a period from May to October 1999. These are large, stable filaments, traditionally classified as quiescent or intermediate \cite{engvold1998}, rather than unstable short-lived ``active region filaments''. However, while the majority of the filaments are located between active regions, our selection criteria do include a small number of filaments forming within active regions. All filaments in the sample were observed for at least 4 consecutive days, but the majority had lifetimes much longer than that.

\begin{figure}
\begin{center}
\leavevmode
\epsfig{file=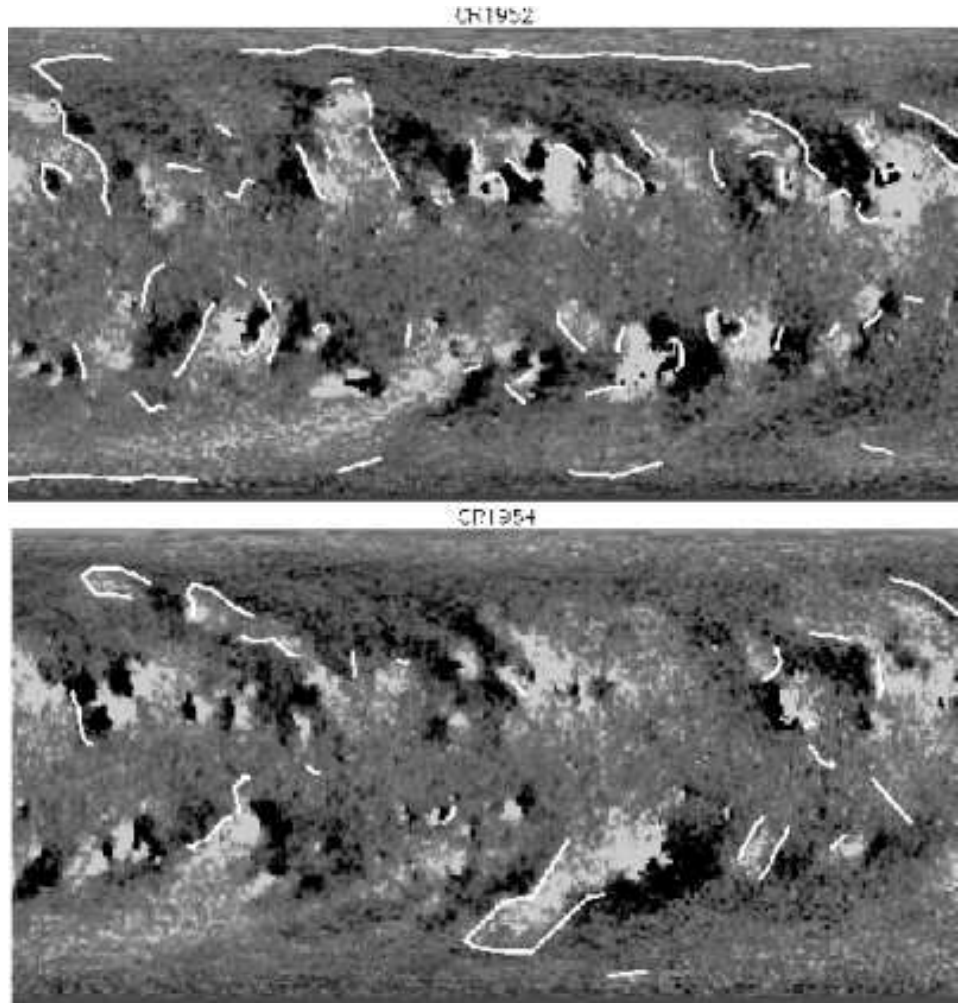,width=1.0\textwidth,clip=}
\caption{Observed filament locations overlayed on Kitt Peak synoptic magnetograms for two selected Carrington rotations, CR1952 and CR1954. On the magnetograms white represents positive flux and black negative flux.}
\label{fig:fils1952}
\end{center}
\end{figure}

The filament locations have been overlayed on synoptic magnetograms from the US National Solar Observatory, Kitt Peak, for the corresponding Carrington rotations CR1949 to CR1954. For example, Figure \ref{fig:fils1952} shows the filament locations for CR1952 and CR1954. Thick white lines indicate the filament locations, and the background shading shows radial magnetic field $B_r$ on the solar surface, with horizontal coordinate longitude and vertical coordinate $\sin\left(\textrm{latitude}\right)$. The filament locations span a range of latitudes, from the equatorial regions to the polar crowns ($60^\circ$ -- $70^\circ$ latitude). Notice how all of the filaments overlie PILs in the photospheric magnetic field. The chiralities of the filaments have been determined by the technique described in Section \ref{sec:chirality}, using the H$\alpha$ images.

\subsection{Filament Chirality} \label{sec:chirality}
%............................................................................

\begin{figure}
\begin{center}
\leavevmode
\epsfig{file=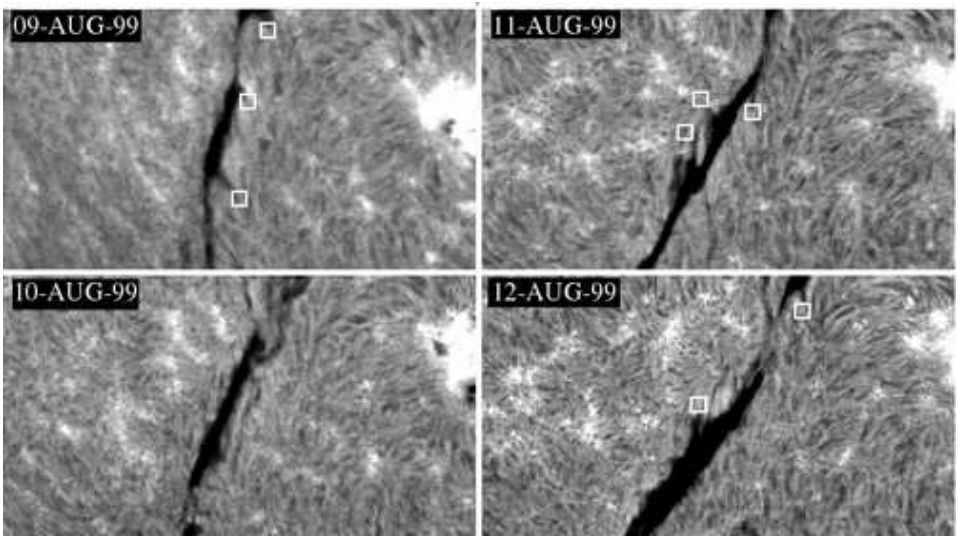,width=1.0\textwidth,clip=}
\caption{Example sequence of H$\alpha$ observations for a single filament, taken on four consecutive days from BBSO full-disk H$\alpha$ images. White squares highlight barb locations identified on the images. From the barb directions this filament is clearly seen to be sinistral.}
\label{fig:obshalpha}
\end{center}
\end{figure}

Each individual filament was followed for up to 7 daily observations about central-meridian passage. An example sequence is shown in Figure \ref{fig:obshalpha} for one particular filament. On each day the angle of view is altered by the Sun's rotation, and a number of barbs may be observed. The orientation of these barbs with respect to the main axis of the filament may be used to determine the filament's chirality, since filaments with right-bearing barbs are dextral, and those with left-bearing barbs are sinistral \cite{martin1998}.

The filament in Figure \ref{fig:obshalpha} is clearly sinistral as all of the observed barbs, shown by white squares, are left-bearing. However, the classification is not always so straightforward, as some filaments appear to have barbs of both chiralities \cite{pevtsov2003}. This may be a result of short-lived perturbations in the local filament structure, or may simply be due to the viewing angle as the filament rotates with the Sun.

To minimise subjectivity in the classification, the following statistical technique was used. For each filament, the number $n_\textrm{d}$ of dextral barbs and the number $n_\textrm{s}$ of sinistral barbs was determined, over all available H$\alpha$ images for that filament. A $t$-test (details in Appendix \ref{sec:stats}) was used to determine whether the filament chirality was significant, based on the observed number of barbs of each type.

\begin{figure}
\begin{center}
\leavevmode
\epsfig{file=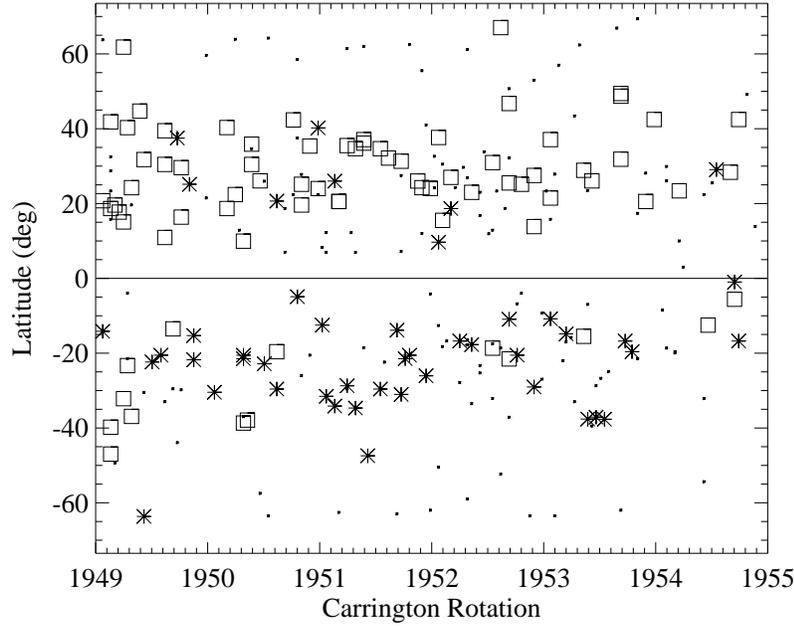,width=0.9\textwidth}
\caption{The 255 observed filaments and their chirality on a time-latitude plot. Squares are dextral filaments and asterisks are sinistral filaments. Small dots indicate filaments where the chirality could not be unambiguously determined from the available H$\alpha$ data.}
\label{fig:latrotbw}
\end{center}
\end{figure}

With a critical value of $t=1.5$ in the statistical test, 99 of the 255 observed filaments could be classified as either dextral or sinistal from the available observations. Of these 64 were dextral and 35 sinistral. Investigation revealed a further 24 filaments to have identifiable chirality, but not enough barbs to pass the $t$-test at this level. Thus the full data set contains 123 filaments of known chirality and 132 of unknown chirality. The classification of each observed filament is shown in Figure \ref{fig:latrotbw}, along with its latitude and date of observation.

Notice that there are a number of exceptions to the hemispheric pattern. Among the filaments of known chirality, 88.7\% follow the hemispheric pattern in the Northern hemisphere, while in the Southern hemisphere 73.1\% follow the pattern. This result is in line with those of \inlinecite{martin1994} and \inlinecite{pevtsov2003}. The aim of this project is to compare directly the chirality of filaments in our sample with the simulated coronal magnetic field. To allow such a comparison we need to model the photospheric field accurately over the observed period, and this is described next.

\section{Surface Model} \label{sec:model}
%----------------------------------------------------------------------------

Accurate simulation of the photospheric magnetic field is a key part to the project, providing the lower boundary condition that drives the 3D coronal field in our model. In this section the flux transport model is described, while the technique for including newly-emerging flux is given in Section \ref{sec:flux}.

\subsection{Flux Transport Equations}
%............................................................................

The transport of magnetic flux on the surface of the Sun has been well studied over a number of years and a standard model has emerged (\opencite{sheeley2005} and references therein). Concentrated flux regions originating at active latitudes are spread over the solar surface by convective motions and sheared by differential rotation. It is now well accepted that there is also a poleward bulk flow, or meridional circulation. These three effects may be incorporated in an advection-diffusion equation for the normal magnetic field on the solar surface \cite{leighton1964,devore1984,wang1989b}. In this paper we write the magnetic field as $\mathbf{B} = \nabla \times \mathbf{A}$ and solve for the vector potential $\mathbf{A}$. The evolution of $B_r$ on the solar surface is then specified in spherical polar coordinates $(r,\theta,\phi)$ by the following pair of equations:
\begin{eqnarray}
\pdiff{A_\theta}{t} &=& u_\phi B_r - \frac{D}{R_\odot \sin\theta}\pdiff{B_r}{\phi},\label{eqn:atheta}\\
\pdiff{A_\phi}{t} &=& -u_\theta B_r - \frac{D}{R_\odot}\pdiff{B_r}{\theta}.\label{eqn:aphi}
\end{eqnarray}
Here $D$ is a diffusion constant that models the random walk of magnetic flux due to the changing supergranular convection pattern (\opencite{leighton1964}; an alternative is the flux-dispersal model of \opencite{schrijver2001}). For our simulations a value of $D=450 \kmsqpsec$ was used. The differential rotation velocity $u_\phi=\Omega(\theta)R_\odot\sin\theta$ is well determined from observations. We use the \inlinecite{snodgrass1983} formula for the angular velocity which, in the Carrington frame, gives
\begin{equation}
\Omega(\theta) = 0.18 - 2.3 \cos^2\theta - 1.62 \cos^4\theta \quad\textrm{deg day}^{-1}.\label{eqn:diffrot}
\end{equation}
The meridional flow is an order of magnitude weaker than the differential rotation so it is observationally less well known \cite{hathaway2005}. As in \inlinecite{mackay4}, we use the profile
\begin{equation} 
u_\theta(\theta) = C\cos\left[\frac{\pi\left(\theta_\textrm{max} + \theta_\textrm{min} - 2\theta \right)}{2\left(\theta_\textrm{max}-\theta_\textrm{min}\right)} \right]\cos\theta,
\end{equation}
where $C$ is chosen to give a maximum flow at mid-latitudes of $16 \mpsec$. In addition to the advection and diffusion terms, a source term is added to Equations \eqref{atheta} and \eqref{aphi} to represent newly-emerging active regions. This allows us to simulate the photospheric evolution continuously for many months.

\subsection{Numerical Method}
%............................................................................

To compute the evolution of $B_r$ on the solar surface the flux transport equations for $A_\theta$ and $A_\phi$ are solved numerically in a two dimensional domain covering both hemispheres, where $10^\circ \leq \theta \leq 170^\circ$ and $0^\circ \leq \phi \leq 360^\circ$. For ease of computation, transformed coordinates $(x,y)$ are used, where
\begin{equation}
x=\phi, \quad y = -\ln\left(\tan\left( \theta/2\right) \right).
\end{equation}
The number of grid points in each direction is $(361,281)$, giving a resolution of $1^\circ$ at the equator. At each stage terms on the right-hand sides of Equations \eqref{atheta} and \eqref{aphi} are computed using finite-differences, before advancing in time with a time step $dt = 60\s$. Spatial derivatives are obtained to second-order accuracy by defining variables on a staggered grid, where $\mathbf{A}$ and $\mathbf{j}$ are defined on cell ribs and $\mathbf{B}$ on the cell faces.

Boundary conditions are applied to $B_r$ and are simply periodic in the $x$-direction and closed in the $y$-direction. The initial condition $B_r(x,y)$ is derived from observed magnetograms as described in Section \ref{sec:mag}.

\subsection{Magnetogram Data and Initial Surface Distribution} \label{sec:mag}
%............................................................................

As observational input to the simulation we use readily available synoptic magnetogram data from the US National Solar Observatory, Kitt Peak. The resolution of this data is sufficient for the present study as we are interested in the large-scale global magnetic field, and in quiescent filaments which may span tens of degrees on the Sun. For the surface simulation the magnetograms are used in two ways: firstly to obtain an initial condition for the simulation, and secondly to determine where new large-scale flux has emerged (Section \ref{sec:flux}).

\begin{figure}
\begin{center}
\leavevmode
\epsfig{file=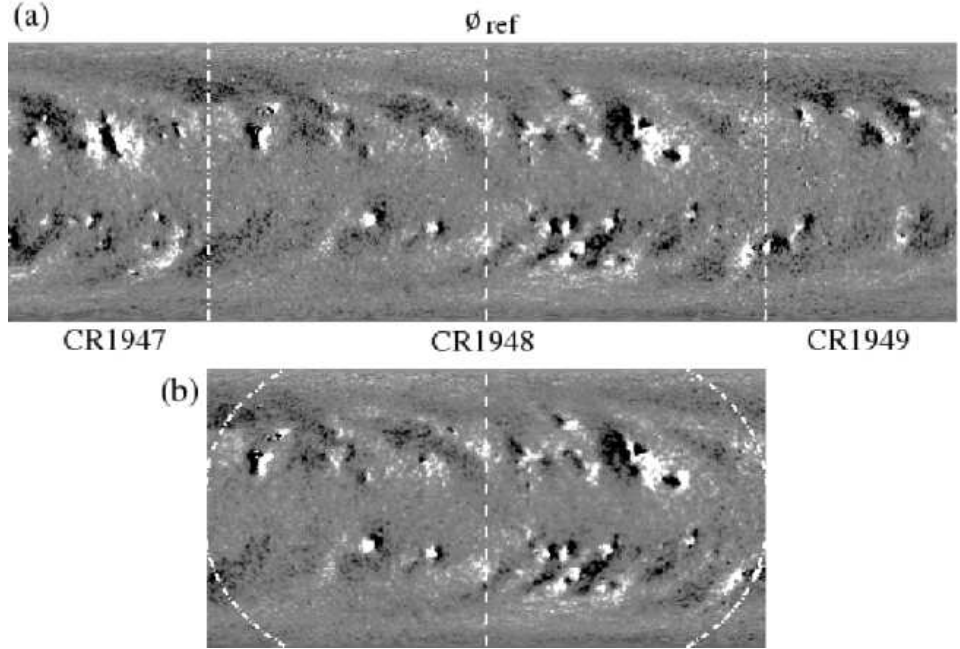,width=1.0\textwidth}
\caption{Correction of synoptic magnetograms for differential rotation, producing the initial condition for the simulation, starting on day 106. Three consecutive maps (a) are used to produce a single corrected map (b), applying the formula \eqref{correct}. The dashed line $\phi=\phi_\textrm{ref}$ shows the reference longitude for day 106, and remains unrotated. The dot-dashed lines show the effect of the rotation on the boundaries of the original CR1948 map.}
\label{fig:diffcorrect}
\end{center}
\end{figure}

Each normal component magnetogram is taken in daily strips over a 27 day period (a single Carrington rotation). This creates a problem when a simultaneous map is needed at all longitudes, as is required for the initial condition in the global simulation. \inlinecite{mccloughan2002} deal with the problem by using a synoptic evolution equation, rather than the flux transport equation itself. We take an alternative approach and produce an initial configuration directly from the magnetogram. A simultaneous magnetic map of the photosphere is created by correcting the observed magnetogram for differential rotation, as illustrated in Figure \ref{fig:diffcorrect} which shows the creation of the initial condition for our simulation. The effects of diffusion and meridional flow are neglected as they are much weaker over a timescale of 13 days, the time between either the left or right edges of the magnetogram and the central reference longitude $\phi_\textrm{ref}$. This is the longitude viewed at central meridian on the start day of our simulation. 

The required correction is dependent on longitude in the magnetogram, which effectively corresponds to time over the 27-day period. In Figure \ref{fig:diffcorrect} we have chosen the reference longitude $\phi_\textrm{ref}=180^\circ$, which was observed mid-way through the 27 days (corresponding to day of year 106 in this case). If the whole map is to represent this single day, then points to the right of $\phi_\textrm{ref}$ must have extra rotation applied, as their central-meridian passage was earlier than $\phi_\textrm{ref}$. Accordingly, points to the left of $\phi_\textrm{ref}$ must have rotation removed. If the original point is at longitude $\phi_0$ and colatitude $\theta$, then the new longitude in our simultaneous map will be
\begin{equation}
\phi_\textrm{n}(\theta) = \phi_\textrm{ref} + \frac{\Omega(\theta) + \Omega_0}{\Omega_0}(\phi_0 - \phi_\textrm{ref}). \label{eqn:correct}
\end{equation}
Here $\Omega(\theta)$ is the differential rotation profile \eqref{diffrot} in the Carrington frame, and $\Omega_0=13.2\,\,\textrm{deg day}^{-1}$ is the rotation rate of the Carrington frame with respect to an observer on Earth. Notice that all points remain at the same colatitude during this operation. The amount of rotation depends on their colatitude, as is clearly seen by the change in shape of the original map boundaries in Figure \ref{fig:diffcorrect}. It can also be clearly seen in Figure \ref{fig:diffcorrect} that we must also make use of the two neighbouring magnetograms to produce a single map using this operation. Thus for our initial condition based on the mid-way point of Carrington rotation CR1948, we must use magnetograms for CR1947, CR1948, and CR1949.

While this technique only gives an approximate map of what would have been on the surface on the start day of our simulation, the simulation results (Section \ref{sec:results}) show that it is sufficiently accurate to reproduce the surface field with future evolution.

\section{Emerging Flux} \label{sec:flux}
%----------------------------------------------------------------------------

In this section we detail the method for selecting newly-emerged active regions and inserting them into the simulation. Over the 6-month timescale of our simulation, a significant amount of magnetic flux will be added by new active regions, driven by the solar dynamo \cite{charbonneau2005}. \inlinecite{schrijver1994} estimate a rate of flux emergence from bipolar regions of $6.2\times 10^{21}\mxday$ near solar maximum. They find that enough flux emerges in these active regions to replenish the entire photospheric flux in about $4$ months. For our 6-month simulation, we must clearly include newly emerging active regions in order to reproduce the surface magnetic field, and subsequently the coronal field, with any realism.

Solar active regions are thought to result from the buoyant rise of sub-photospheric magnetic flux tubes which break through the surface in an $\Omega$ loop, producing the characteristic bipolar magnetic configuration \cite{parker1955,babcock1961}. These magnetic bipoles are oriented in an approximately east-west direction, although the bipole axis is typically tilted so that the leading polarity is nearer to the equator \cite{hale1919,wang1989}. The vast majority of bipoles in the Northern hemisphere have the same leading polarity, and those in the Southern hemisphere have the opposite polarity \cite{hale1925}. With the global reversal of the Sun's polar magnetic field every 11 years, all the polarities reverse.

Since the model does not include a description of the magnetic field inside the Sun, we cannot predict when and where magnetic flux will emerge. As the ultimate aim is to reproduce observed coronal features, it would be inappropriate to insert new active regions at random locations and times during the evolution, as has been used in some previous surface flux transport models \cite{vanballegooijen1998, mackay5b, baumann2004, mackay2004}. Instead the observed magnetograms throughout the simulation period are analysed to determine where and when new flux has emerged \cite{sheeley1985}. Magnetic flux is then inserted into the simulation in the form of magnetic bipoles with properties matching those of the observed active regions. Of course, such a technique based on the synoptic magnetograms will only pick up the larger emerging flux regions, but we believe that this is sufficient to produce the large-scale field over long time-scales, as is required in this study.

\subsection{New Flux Regions} \label{sec:newflux}
%............................................................................

\begin{figure}
\begin{center}
\leavevmode
\epsfig{file=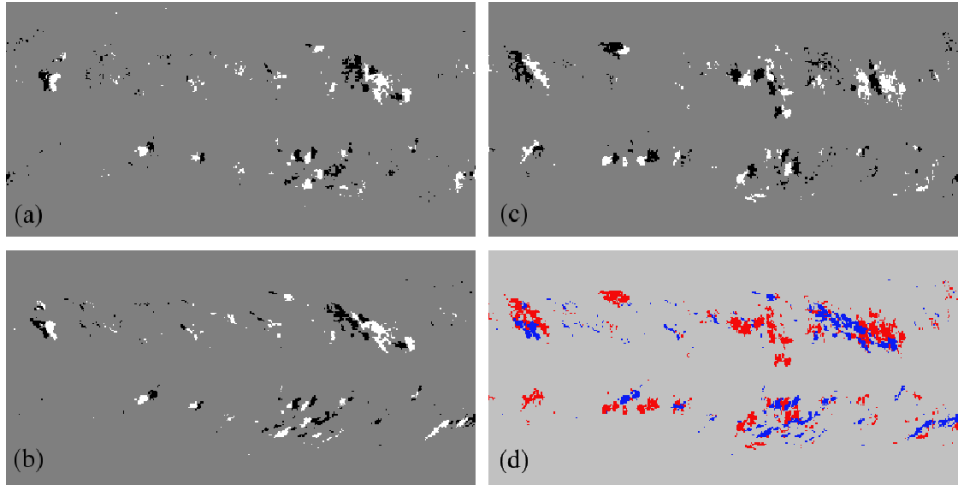,width=1.0\textwidth,clip=}
\caption{Stages in processing the synoptic magnetograms for the detection of new bipolar regions: (a) observed map for CR1948, (b) result of applying 27.27 days of differential rotation, (c) observed map for CR1949, and (d) absolute difference map. In (a), (b) and (c), white represents magnetic flux above $50\gauss$ and black represents flux below $-50\gauss$. In (d) red indicates new areas of flux and blue indicates old areas.}
\label{fig:detection}
\end{center}
\end{figure}

Locations of new flux emergence are found by comparing successive synoptic magnetograms. Some care is necessary in doing so because the field will have evolved in the intervening 27 days. The comparison is made easier by applying the correct amount of differential rotation to the earlier magnetogram, as in Section \ref{sec:mag}. Figure \ref{fig:detection} illustrates the procedure for a particular case. The earlier observed magnetogram (CR1948) is shown in Figure \ref{fig:detection}(a), and it has been corrected for differential rotation with reference longitude $\phi_\textrm{ref}=0^\circ$ (corresponding to the final day of that rotation). The resulting magnetogram is shown in Figure \ref{fig:detection}(b). This is the magnetogram which is compared with the observed magnetogram for the next rotation (CR1949), shown in Figure \ref{fig:detection}(c). In Figures \ref{fig:detection}(a) to (c), white areas represent magnetic flux above $50\gauss$ and black areas represent flux below $-50\gauss$.

A semi-automated procedure has been developed to determine the locations of new flux, based on some simple criteria. This is found effective in reducing the time taken for a manual search. The stages in this procedure are as follows:
\begin{enumerate}
\item{Determine the boundary coordinates of all regions of flux on the later observed magnetogram. These regions are defined as continuous pixel groups with radial field $|B_r|$ greater than a threshold of $50\gauss$.}
\item{Compute an absolute difference map between the later observed magnetogram and the rotated earlier magnetogram. An example is shown in Figure \ref{fig:detection}(d). Here red areas represent flux (of either sign) present only in the later observed map, and blue areas represent flux present only in the rotated earlier map.}
\item{Select all regions from first stage where there is a majority of red flux on the difference map. These represent new regions.}
\item{Flag any anomalous regions for manual attention. These include regions where there is mostly only one polarity of flux, regions that do not appear to follow the Hale polarity law, and also any regions not considered to be new which have very high peak strength. All probable new regions which have not been flagged are then classified as ``definitely new''.}
\item{The flagged regions are included or excluded manually, and the user can select any further regions as desired, in order to improve the accuracy of the simulation.}
\end{enumerate}
The final step has been particularly useful in the case of large activity complexes, which may consist of multiple bipolar regions in close proximity. On some occasions it has been necessary to add in a new bipole in existing active regions where more flux appears to have emerged. A total of 119 bipolar regions were determined in the 6 Carrington rotations CR1949 to CR1954 inclusive. Of these, 97 were detected by the automated program, the remainder being added manually.

\begin{figure}
\begin{center}
\leavevmode
\hbox{
\epsfig{file=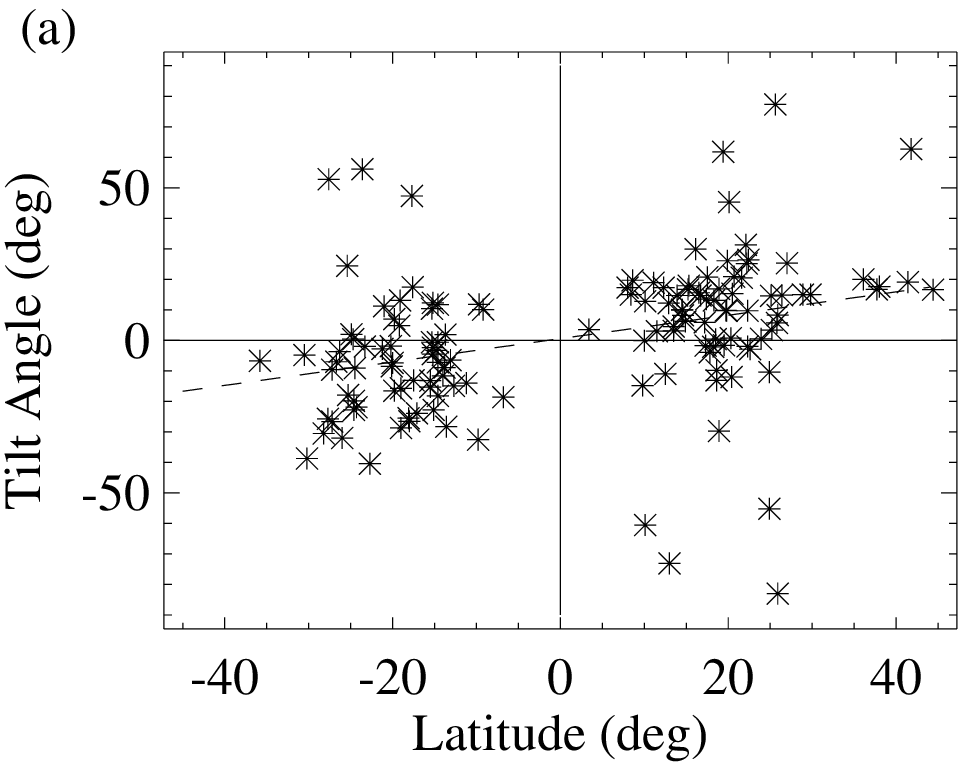,width=0.5\textwidth,clip=}
\epsfig{file=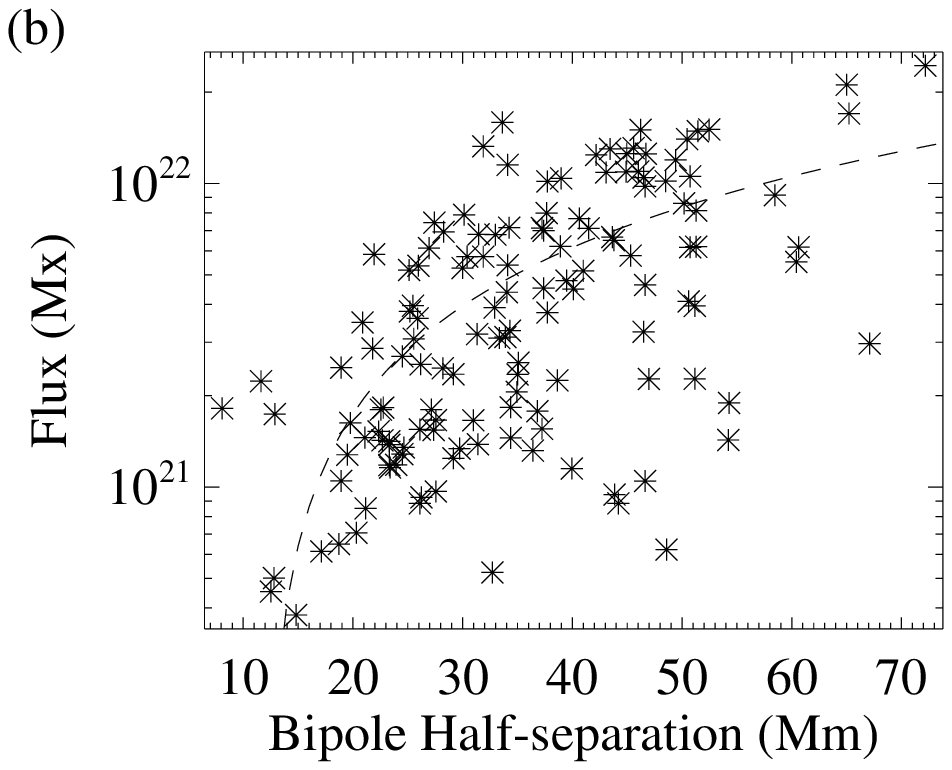,width=0.5\textwidth,clip=}}
\caption{Summary of properties of the 119 bipolar regions: (a) tilt angle $\delta$ against bipole centre latitude, and (b) magnetic flux against half-separation distance $\rho$ between centres of each polarity. In both plots the dashed line is a linear least-squares fit.}
\label{fig:tiltflux}
\end{center}
\end{figure}

Having located the bipolar regions where new flux has emerged, the following properties of these regions are measured: day of central-meridian passage, coordinates of bipole centre, tilt angle, half-separation distance between peaks, and flux. The locations of the bipole peaks were determined by taking the centroid of flux of each polarity. Figures \ref{fig:tiltflux}(a) and (b) summarize the data for the 119 observed new regions. Figure \ref{fig:tiltflux}(a) shows the typical relation between bipole tilt angle and latitude of emergence known as Joy's Law \cite{hale1919}. We find that the average tilt angle, from a least-squares fit, varies as $0.38\times \textrm{latitude}$, but with a large spread of data about the regression line as seen in Figure \ref{fig:tiltflux}(a). This is in good agreement with \inlinecite{wang1989}, who found a slope of about $0.4$, and a similarly large spread. In Figure \ref{fig:tiltflux}(b) we see the expected positive correlation between total magnetic flux and active region size \cite{schrijver1994}.

Our 119 emerging regions contribute a total flux $\left|\Phi_+\right| + \left|\Phi_-\right|$ of $1.18 \times 10^{24}\mx$ over the 177-day simulation. This agrees with the estimate of \inlinecite{schrijver1994} for flux emergence from bipolar regions near solar maximum. It is significantly higher than the total flux present in the photosphere at any one time, which at the start of our simulation is $5.09\times 10^{23}\mx$. Emerging flux is clearly vital to the photospheric field evolution over this 6-month period, and must clearly be included in our simulations.

\subsection{Bipole Description}
%............................................................................

Emerging flux is included in our simulation by the insertion of magnetic bipoles with properties matching those of the observed regions. The magnetic bipoles all take the same functional form, as used in \inlinecite{mackay2001} and \inlinecite{mackay4}. This is a three-dimensional field, allowing not only the insertion of the bipole at the photospheric level, but also in the corona. In the corona this bipole field may be non-potential, representing the helicity contained in newly-emerging regions on the Sun. In our computational $(x,y,z)$ coordinates a bipole is given by the vector potential
\begin{eqnarray}
A_x &=& \beta B_0 \expt{0.5}z\expt{-2\xi},\\
A_y &=& B_0\expt{0.5}\rho \expt{-\xi},\\
A_z &=& -\beta B_0 \expt{0.5}x\expt{-2\xi},
\end{eqnarray}
where
\begin{equation}
\xi = \frac{\left(x^2+z^2\right)/2 + y^2}{\rho^2},
\end{equation}
and the parameter $\rho$ is the half-separation distance between the peaks of the two polarities. This is for a bipole with zero tilt angle; for non-zero tilt angle $\delta$ the field is simply rotated. The twist parameter $\beta$ controls the helicity of the bipole. In the present paper only the photospheric footprint of this field will be required, which does not depend on $\beta$, but the helicity will be important in the future coronal simulations. 

\begin{figure}
\begin{center}
\leavevmode
\hbox{
\epsfig{file=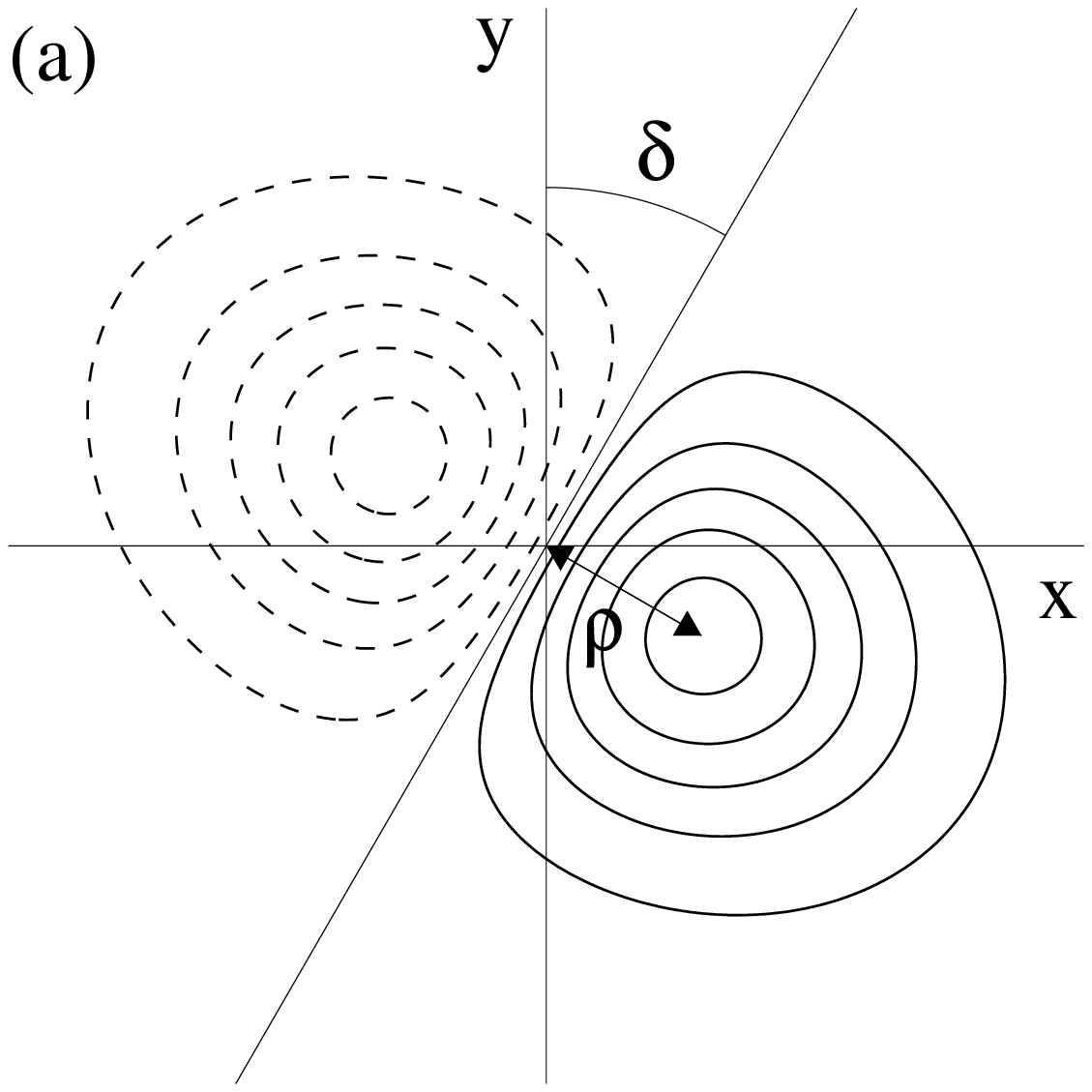,width=0.4\textwidth,clip=}
}
\hbox{
\epsfig{file=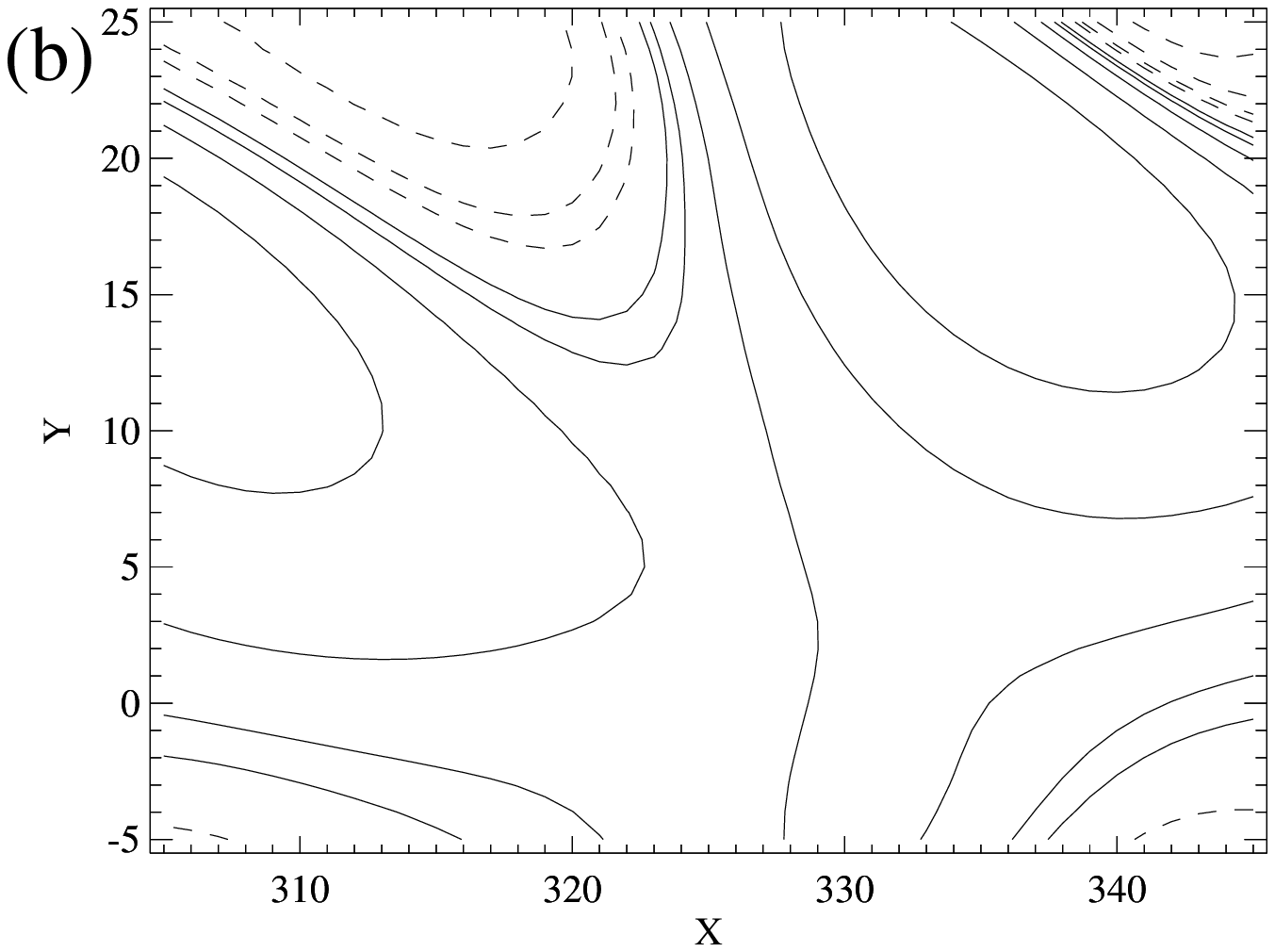,width=0.5\textwidth,clip=}
\epsfig{file=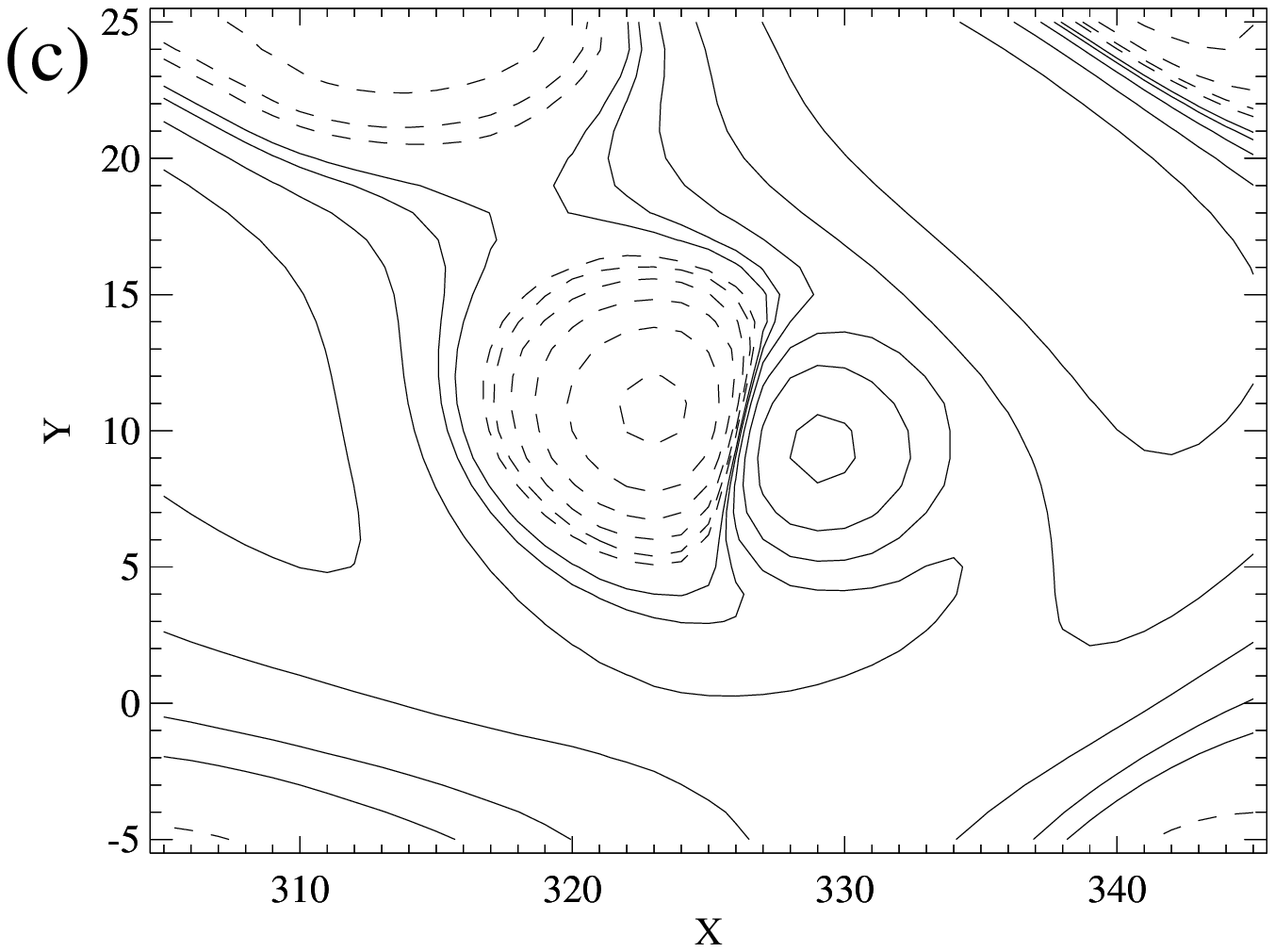,width=0.5\textwidth,clip=}
}
\caption{Illustration (a) of the bipole given by Equation \eqref{brbipole}, with tilt angle $\delta$ and half-separation $\rho$. An example insertion from the simulation is seen ``before and after'' on (b) day 250 and (c) day 251. All three plots show magnetic field strength $B_r$, with solid contours being positive flux and dashed contours negative flux.}
\label{fig:bipole}
\end{center}
\end{figure}

For the present we set $z=0$ and disregard the $A_z$ component. This gives the radial surface magnetic field
\begin{equation}
B_r = -B_0\expt{0.5}\frac{x}{\rho}\exp\left[-\left(\frac{x^2/2+y^2}{\rho^2}\right)\right],
\label{eqn:brbipole}
\end{equation} 
which is illustrated in Figure \ref{fig:bipole}(a) for a non-zero tilt angle $\delta$ corresponding to a rotation of the coordinates $(x,y)$ in Equation \eqref{brbipole}. Note that in Equation \eqref{brbipole} $B_r$ is independent of the twist parameter $\beta$, so that changing $\beta$ alters the coronal field but gives the same photospheric field. Finally, the observations give the total flux $\Phi$ of the bipole rather than the peak field $B_0$, but for the model bipole the two are related by
\begin{equation}
B_0 = \frac{\Phi}{\sqrt{\pi}\rho^2\expt{0.5}}.
\end{equation}

\subsection{Insertion of New Regions}
%............................................................................

New bipoles are inserted by adding their corresponding vector potential, as described in the previous section. There are two complicating factors; firstly the exact day of emergence is unknown, as some bipoles may emerge on the unobserved side of the Sun, and secondly there may be pre-existing flux at the emergence location.

Because the observed synoptic magnetograms have a time cadence of only 27 days, there is considerable uncertainty about the exact emergence date of new flux regions. In particular it is unlikely that a region appeared on the exact day it was observed at central-meridian, and it may well have emerged on the unobserved far-side of the Sun. In view of this, and to allow time for coronal structures to develop before the date of filament observation at central-meridian passage, we have emerged all bipoles 7 days before their date of observation. To ensure accuracy we have ``evolved'' the parameters of each bipole back in time by 7 days (see Appendix \ref{sec:backward} for more details). Seven days is an arbitrary choice, but we have varied the number of days and found no critical effect. This aspect of our model could be improved in future, either by more accurate modelling of the development of individual active regions, or by using newly-developed helioseismic far-side imaging techniques \cite{lindsey2000}. Note however that such techniques are not yet sufficiently developed for use in this present work.

Now consider how a new bipole will be inserted in a region with a considerable pre-existing magnetic field. High coronal field lines previously connected to the photosphere in that region find themselves with nowhere to connect down after the new bipole is inserted. To avoid any associated problems in the three-dimensional simulation we have adopted a technique of ``sweeping'' to move any earlier strong field out of the region of insertion. This is not without some physical basis, as one might expect a newly emerging flux tube to distort the previous coronal field out of its path \cite{yokoyama1996,krall1998}.

While the sweeping takes place the simulation time is frozen, so that the bipole insertion is still instantaneous. The ordinary flows are turned off, and a new outward velocity of the form $v_s=\exp(-As)$ is applied in the photospheric and coronal region surrounding the bipole centre, where the coordinate $s$ represents distance from the bipole centre. The constant $A$ is chosen so as to give zero flow at the edge of the insertion region. This sweeping velocity advects the field out of the insertion region, and sweeping stops once the field near the bipole centre reaches a suitably low level. We have found that a threshold of $0.05B_0$ works well, where $B_0$ is the maximum magnetic field of the new bipole.

Figures \ref{fig:bipole}(b) and (c) show the insertion of a new region on day 251 of the simulation. There is a considerable pre-existing field on day 250 (Figure \ref{fig:bipole}b), and the bipole emerges in what was largely a region of positive flux (solid contours), but with an encroaching region of negative flux (dashed contours) at higher latitude. Considerable sweeping was needed to reduce the field strength in the insertion region. This has resulted in the pre-existing negative flux being pushed up to higher latitude to make way for the new bipole in Figure \ref{fig:bipole}(c). The bipole insertion procedure in 3D will be described and illustrated when the 3D coronal field simulations are discussed in a following paper. This will demonstrate how our newly inserted bipoles reconnect with the overlying coronal field to produce a plausible magnetic configuration.

By the method outlined in this section we have successfully incorporated the insertion of newly-emerging bipoles into the photospheric simulation, allowing continuous evolution over many months. Furthermore, the bipoles are three-dimensional and non-potential in nature. This represents a significant advance over previous models, and in the subsequent 3D simulations will allow us to study the build-up of magnetic helicity and shear in the coronal field over long periods, driven by the emergence and evolution of active regions.

\section{Surface Simulation Results} \label{sec:results}
%----------------------------------------------------------------------------

In this section we illustrate the newly developed surface simulations over a particular 6 month period from April to October 1999. This will form the continuous photospheric boundary condition for the global coronal magnetic field simulations to be described in a following paper. The simulation starts from day-of-year 106, mid-way through Carrington rotation CR1948. It then runs continuously for 177 days until day 283 at the end of CR1954. Two different simulation runs are illustrated: one where no emerging flux is inserted during the simulation, and one where the 119 bipolar regions described in Section \ref{sec:newflux} are inserted at the relevant days. In both cases the initial condition is the corrected magnetogram for day 106 (Figure \ref{fig:diffcorrect}b).

\begin{figure}
\begin{center}
\leavevmode
\epsfig{file=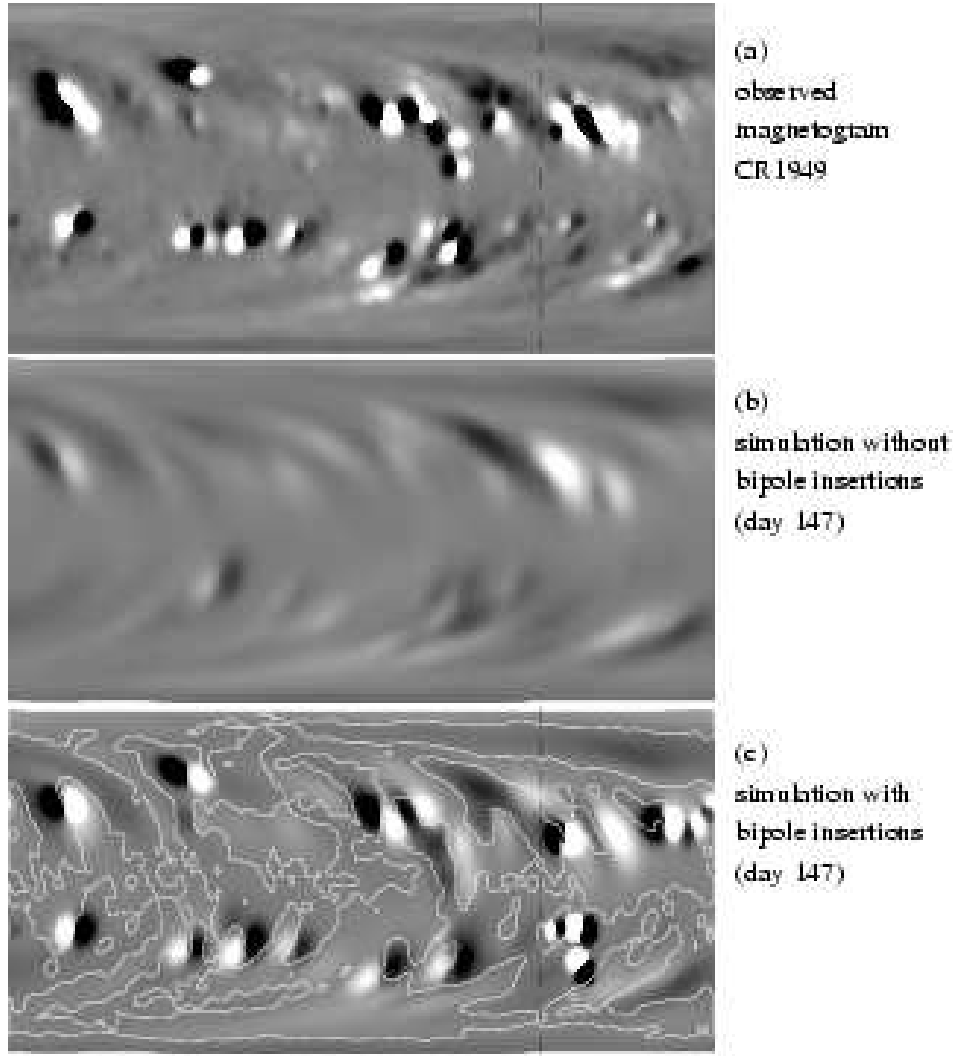,width=1.0\textwidth,clip=}
\caption{Comparison of simulation results at day 147: (a) observed magnetogram corrected for differential rotation, (b) simulation with no bipole insertions, and (c) simulation with bipole insertions. In all three plots, white indicates positive flux and black negative, with a saturation level of $\pm 50 \gauss$. On the simulated magnetogram (c), the zero contour from the \emph{observed} magnetogram (a) is overlayed in white. To the right of the black dashed line, new regions from the next rotation will appear in (c) but not in (a), because we insert them 7 days before observation at central meridian.}
\label{fig:compare147}
\end{center}
\end{figure}

\begin{figure}
\begin{center}
\leavevmode
\epsfig{file=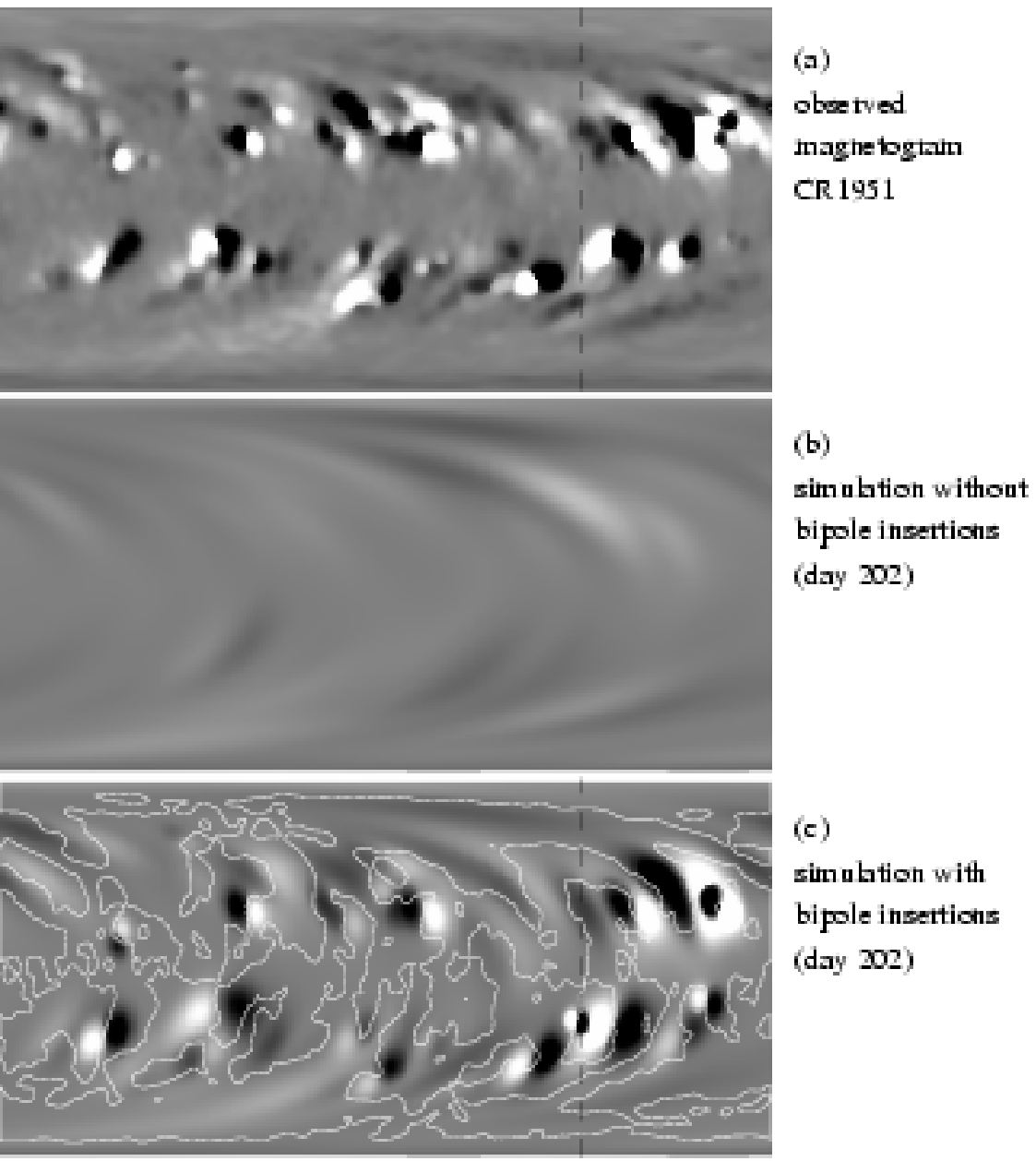,width=1.0\textwidth,clip=}
\caption{Comparison of simulation results at day 202: (a) observed magnetogram corrected for differential rotation, (b) simulation with no bipole insertions, and (c) simulation with bipole insertions. In all three plots, white indicates positive flux and black negative, with a saturation level of $\pm 50 \gauss$. On the simulated magnetogram (c), the zero contour from the \emph{observed} magnetogram (a) is overlayed in white. To the right of the black dashed line, new regions from the next rotation will appear in (c) but not in (a), because we insert them 7 days before observation at central meridian.}
\label{fig:compare202}
\end{center}
\end{figure}

\begin{figure}
\begin{center}
\leavevmode
\epsfig{file=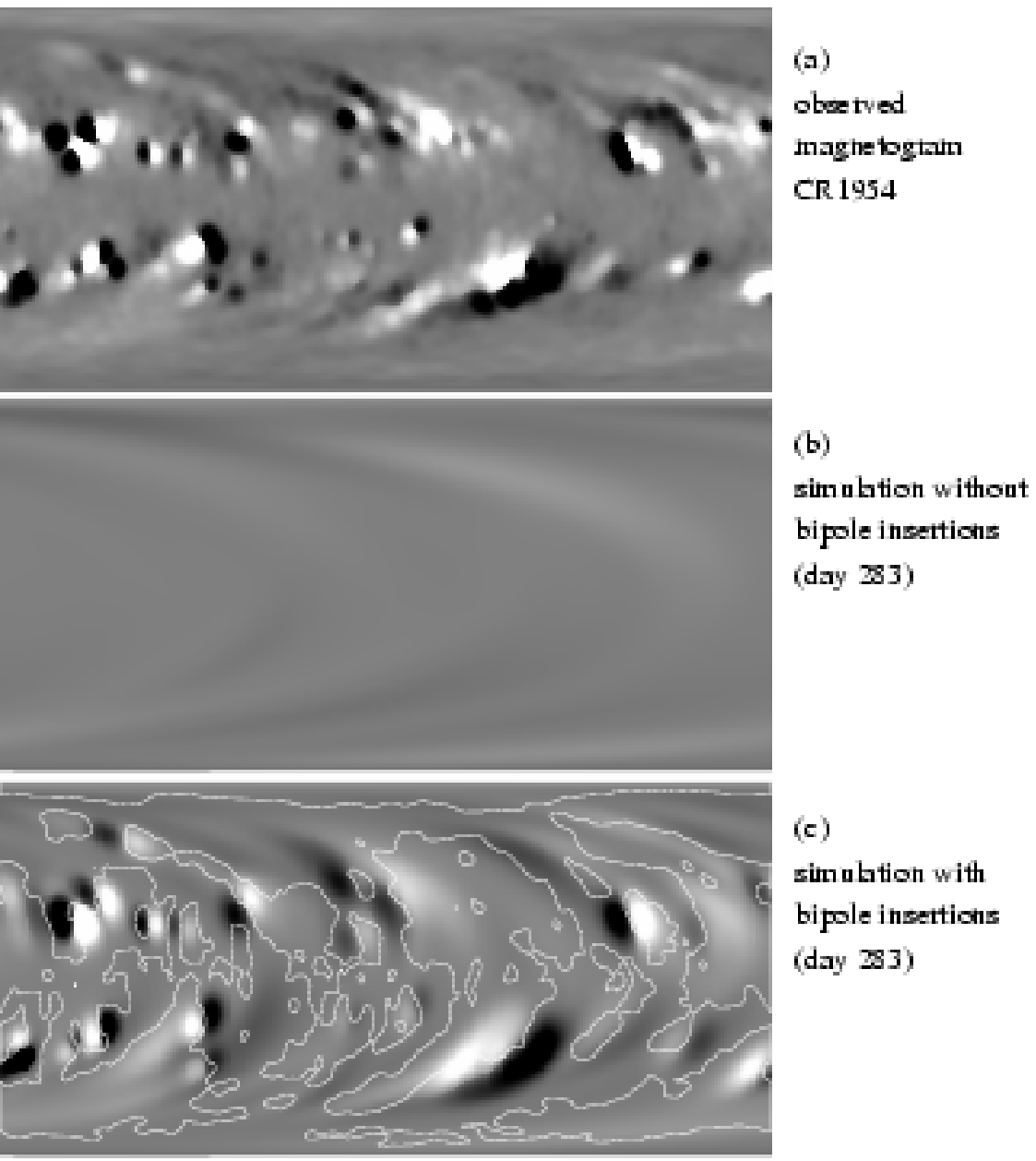,width=1.0\textwidth,clip=}
\caption{Comparison of simulation results at the end of the simulation, on day 283: (a) observed magnetogram corrected for differential rotation, (b) simulation with no bipole insertions, and (c) simulation with bipole insertions. In all three plots, white indicates positive flux and black negative, with a saturation level of $\pm 50 \gauss$. On the simulated magnetogram (c), the zero contour from the \emph{observed} magnetogram (a) is overlayed in white.}
\label{fig:compare283}
\end{center}
\end{figure}

Figures \ref{fig:compare147} to \ref{fig:compare283} illustrate the results at different stages in the simulation. In each figure, (a) is the observed synoptic magnetogram, (b) is the simulation run with no bipole insertions, and (c) is the simulation run with bipole insertions. All three of these figures correspond to the final day of a particular Carrington rotation, rotations CR1949, CR1951, and CR1954. The observed magnetogram has been corrected for differential rotation (Section \ref{sec:mag}) with reference longitude $\phi_\textrm{ref}=0^\circ$, giving an approximation to the instantaneous field of the Sun on the final day of the given Carrington rotation. This is to facilitate comparison with the simulated field, and the observed magnetogram has also been smoothed for easier comparison. In each plot the horizontal coordinate is longitude and the vertical coordinate is sin(latitude). All plots use the same levels for $B_r$, with a saturation level of $\pm 50 \gauss$. Black areas represents negative flux and white areas positive flux. For each day, the zero contour of $B_r$ from the observed magnetogram (a) is overlayed on the simulated magnetogram (c) for comparison.

After 1 month of evolution the simulation with no bipole insertions (Figure \ref{fig:compare147}b) reproduces the diffuse background field well, particularly nearer to the poles. This magnetic flux results from the dispersal of earlier active regions which have been carried poleward by the meridional circulation. The relatively slow speed of the circulation, at most $16 \mpsec$, ensures that flux emerged in the active region belts since the start of the simulation has not yet reached these higher latitudes. At lower latitudes there are numerous regions of strong flux missing from this simulation; these are the active regions that have emerged since the start of the run.

When emerging regions are included the simulation reproduces the observed field at day 147 well, as evidenced by visual inspection and the overlayed zero contour on Figure \ref{fig:compare147}(c). The extra bipolar regions in the right-most $90^\circ$ of the simulated magnetogram (to the right of the black dashed line) should be ignored in the comparison. These belong to the observed magnetogram for the next rotation, CR1950, and so do not appear in Figure \ref{fig:compare147}(a) which is the observed magnetogram for CR1949. They appear on this day in the simulation because all of our new regions are inserted 7 days before their date of observation at central-meridian passage.

By day 202 (Figure \ref{fig:compare202}) the simulation with no bipole insertions still reproduces the field well near the poles, but it has lost all stronger flux regions at lower latitudes. However, the simulation with emerging flux maintains good accuracy. It is less accurate in complex flux regions, such as in the top right of Figure \ref{fig:compare202}(c), but the large-scale picture is still reasonable. In principle the structure in these regions could be improved by carefully inserting a larger number of small bipoles, but this would be labour intensive. In any case, such regions become smoothed out over time and the small-scale complexity is lost.

The contrast between the two simulations is even more evident on day 283, where the simulation with no bipole insertions (Figure \ref{fig:compare283}b) is now beginning to lose accuracy even near to the poles, with the field very weak. When bipoles are inserted (Figure \ref{fig:compare283}c) the simulation still gives a good reproduction of the large-scale field, containing both strong newly emerged flux and the diffuse remnants of earlier active regions. Therefore it can be seen that our technique for reproducing the surface field is able to maintain accuracy to the observed field.

\section{Conclusion} \label{sec:conclusions}
%----------------------------------------------------------------------------

In this paper we have developed a new technique for simulating the observed photospheric magnetic field on a global scale over many months, without having to reset to observed magnetograms every 27 days. In a follow-up paper these new simulations will be coupled with simulations of the 3D coronal magnetic field, where our motivation is to explain the hemispheric pattern of the axial magnetic-field direction in solar filaments, a property known as filament chirality.

The essential features of the new photospheric magnetic-field simulations are as follows:
\begin{enumerate}
\item{A surface flux transport model, based on synoptic magnetogram data from the real Sun.}
\item{Accuracy is maintained not by resetting to the observed field, but by the emergence of new active regions throughout the course of the simulation.}
\item{These active regions are in the form of magnetic bipoles, with properties chosen to match those of bipolar regions observed in the synoptic magnetograms.}
\end{enumerate}

We do not reset the photospheric field because in the future 3D simulations this would remove memory of previous field-line connectivity in the coronal field. It would prevent the long-term build-up of non-potential magnetic energy and helicity, which are key to the formation of filament magnetic structures. An advantage of the magnetic bipoles we insert is their mathematical form, which allows them to be inserted also into the corona in the 3D simulation. They can then be given a non-zero magnetic helicity, and such active-region helicity is thought to be important for filament chirality \cite{mackay2003,mackay3}.

Our semi-automated technique for finding new bipolar regions has been found to work well, with the program detecting automatically 97 of the 119 regions inserted in the 6-month simulation. The extra regions have been chosen manually; this has largely been necessary for areas with many bipoles in close proximity, such as in activity complexes. With the 119 emerging regions, the simulation is successful in reproducing the large-scale photospheric field, even over a period of 6 months. The accuracy of the simulation is seen clearly when the observed polarity inversion line is overlayed on the simulated magnetogram. Larger areas of positive and negative magnetic flux are well reproduced, as is the shape of the inversion line at most of the observed filament locations.

It is these locations where our simulated 3D magnetic field will be compared with the chirality of observed filaments. In a recent paper, \inlinecite{mackay3} have shown how the observed properties of active regions on the Sun, along with the observed large-scale surface motions, can produce the observed hemispheric pattern. However, they considered a simple pair of bipolar magnetic regions, simulating only a localised region of the solar photospheric and coronal magnetic field. In the present study, we will test their conclusion by comparing global simulations with real filament data, over a 6-month period. In the global simulations, the 3D coronal field will be coupled directly to the continuous evolution of the photospheric field, so the project relies heavily on the accuracy with which we have been able to simulate the observed photospheric field in this paper.

For comparison with the simulation, we will use a filament data set which includes up to 7 daily observations of each of 255 filaments, taken from BBSO H$\alpha$ images over the 6-month period. As described in this paper, we have determined the chiralities of individual observed filaments using a statistical technique based on the orientations of their barbs. The chirality has been unambiguously determined as either dextral or sinistral for 123 filaments, and it is at these locations where we will compare the observed chirality with the direction of skew in the 3D coronal magnetic field simulations.

\acknowledgements
Financial support for ARY and DHM was provided by the UK Science and Technology Facilities Council (STFC). The simulations were performed on the UKMHD parallel computer in St Andrews, funded jointly by SRIF/STFC. We would like to thank S.F. Martin for supplying the original filament data set, based on data from BBSO/NJIT, as well as travel support from the PROM group (NSF grant ATM-0519249). We also thank C. Stone for developing an initial code for determining observed filament chirality, and P.E. Jupp for advice on statistical calculations. Synoptic magnetogram data from NSO/Kitt Peak was produced cooperatively by NSF/NOAO, NASA/GSFC, and NOAA/SEL and made publicly accessible on the World Wide Web.

\appendix
\renewcommand{\theequation}{\thesection\arabic{equation}}
\setcounter{equation}{0}  % reset counter 
\section{Statistical Test for Filament Chirality} \label{sec:stats}
%----------------------------------------------------------------------------

A statistical $t$-test is used to determine filament chirality from classification of individual barbs. Suppose a single filament has $n$ classifiable barbs, giving $n$ observations $x_1,x_2,\ldots,x_n$, where $x_i=+1$ or $-1$ according as the barb is right-bearing or left-bearing respectively. Then the number of dextral barbs is $n_\textrm{d} = \sum_{x_i=1}x_i$, and the number of sinistral barbs is $n_\textrm{s}=n-n_\textrm{d}$. It is reasonable to assume that $n_\textrm{d}$ follows a binomial distribution with parameters $(n,p)$, and we assume $p=0.5$.

The test is based on the statistic
\begin{equation}
t = \frac{n_\textrm{d} - np}{\sqrt{np(1-p)}},
\label{eqn:ttest}
\end{equation}
which should be near to $0$ if neither chirality is significant. The classification scheme is then
\begin{eqnarray*}
t & >&  T \quad \textrm{dextral},\\
t & <&  -T \quad  \textrm{sinistral},\\
|t| & \leq&  T \quad  \textrm{not classified},
\end{eqnarray*}
where $T$ is some chosen threshold. For large enough $n$, $t$ should approximate a normal distribution with mean $n$ and variance $1$. As our sample sizes are relatively small we choose $T=1.5$. For a filament with 9 barbs, Equation \eqref{ttest} shows that this corresponds to a threshold ratio $n_\textrm{d}/n = 0.75$.

\section{Backward Evolution of Bipole Parameters} \label{sec:backward}
%----------------------------------------------------------------------------
\setcounter{equation}{0}  % reset counter 

To insert a bipole 7 days before its date of observation at central-meridian passage, the bipole properties must be ``evolved'' back in time for 7 days, so as to match the observations as closely as possible when it does reach central meridian. The properties in question are the location of the bipole centre $(\theta_\textrm{cen},\phi_\textrm{cen})$, the tilt angle $\delta$, the half-separation $\rho$ between centres of polarity, and the magnetic flux.

The movement of the centres of each polarity may be inferred from the formula \eqref{diffrot} for differential rotation $\Omega(\theta)$, as meridional flow may be effectively neglected over such a timescale. This allows the evolution of the bipole centre, tilt angle, and half-separation to be calculated. Suppose the bipole properties are known at time $t=t_0$. Then if the centre of the leading polarity is at $(\theta_1,\phi_1(t))$, and that of the following polarity is at $(\theta_2,\phi_2(t))$, it can be shown that
\begin{eqnarray}
 \phi_\textrm{cen}(t) &=& \phi_\textrm{cen}(t_0) + \frac{\Omega(\theta_1) + \Omega(\theta_2)}{2}t, \label{eqn:phicenevol} \\
 \theta_\textrm{cen}(t) &=& \theta_\textrm{cen}(t_0) \quad \textrm{for all $t$}, \label{eqn:thetacenevol}\\
 \tan \delta(t) &=& \frac{2\rho(t_0)\sin\delta(t_0)}{2\rho(t_0)\cos\delta(t_0) + R_\odot\sin\theta_\textrm{cen}\Omega_0 t}, \label{eqn:tanevol}\\
 \rho(t)& =& \sqrt{\left( \frac{\rho(t_0)\cos\delta(t_0)}{R_\odot\sin\theta_\textrm{cen}} + \frac{\Omega_0}{2}t \right)^2 R_\odot^2 \sin^2\theta_\textrm{cen}   + \left( \rho(t_0)\sin\delta(t_0)\right)^2}, \label{eqn:rhoevol}\nonumber\\
&&
\end{eqnarray}
where
\begin{equation}
\Omega_0 = \Omega\left(\theta_+ \right) - \Omega\left(\theta_-\right),
\end{equation}
and the colatitudes $\theta_+$ and $\theta_-$ are given by
\begin{equation}
\theta_+ = \theta_\textrm{cen} + \frac{\rho(t_0)\sin\delta(t_0)}{R_\odot} \quad \textrm{and} \quad \theta_- = \theta_\textrm{cen} - \frac{\rho(t_0)\sin\delta(t_0)}{R_\odot}.
\end{equation}

The magnetic flux above the $50\gauss$ threshold will also change, and depends on both advection and diffusion. It has been determined approximately by test simulations running in the forward-time direction. By varying the latitude, tilt-angle, and half-separation, a look-up table was created relating the initial and final flux for different parameters. It was not necessary to include different values of initial flux in this look-up table, as varying the initial flux simply results in a proportional scaling of the flux at all times.

\end{article} 
\end{document}